\newif\ifcomments
\commentstrue 

\documentclass[conference]{IEEEtran}

\usepackage{color}
\usepackage{amssymb}
\usepackage{amsmath}
\usepackage{bm}
\usepackage{graphicx}

\newtheorem{definition}{Definition}
\newtheorem{theorem}{Theorem}
\newtheorem{lemma}[theorem]{Lemma}

\newcommand{\prob}{\textrm{Pr}}

\newcommand{\error}[1]{\epsilon_{#1}}
\newcommand{\shift}[1]{\delta_{#1}}
\newcommand{\ratio}[1]{\gamma_{#1}}

\newcommand{\event}[1]{\text{E}_{#1}}
\newcommand{\bigo}[1]{\mathcal{O}({#1})}

\newcommand{\cN}{{\mathcal{N}}}
\newcommand{\cD}{{\mathcal{D}}}
\newcommand{\bl}{n}
\newcommand{\dfct}{d}

\newcommand{\lowt}{l}
\newcommand{\upt}{u}
\newcommand{\gap}{g}

\newcommand{\test}{T}

\newcommand{\inp}{x}
\newcommand{\outp}{y}

\newcommand{\algoa}{{\bf TGT-BERN-NONA}}
\newcommand{\algob}{{\bf TGT-BERN-ADA}}
\newcommand{\algoc}{{\bf TGT-LIN-NONA}}
\newcommand{\reference}{reference }
\newcommand{\indicator}{indicator }
\newcommand{\critical}{critical}

\newcommand{\dstr}{\rho}
\newcommand{\family}{r}
\newcommand{\person}{j}
\newcommand{\stage}{i}
\newcommand{\pool}{k}

\newcommand{\parta}{\mathcal{P}}
\newcommand{\bin}{\bar{\parta}}
\newcommand{\rfrn}{\mathcal{R}}
\newcommand{\idct}{\mathcal{I}}
\newcommand{\crfrn}{\bm{\rfrn}}
\newcommand{\cidct}{\bm{\idct}}

\newcommand{\bratio}{\ratio{1}}

\newcommand{\bnum}{P}
\newcommand{\rrpt}{R}
\newcommand{\irpt}{I}
\newcommand{\iratio}{\ratio{2}}

\newcommand{\outptu}[3]{\outp^{(#1)}_{#2,#3}}
\newcommand{\rtrsd}[1]{\theta_{#1}}
\newcommand{\rtail}[1]{\Delta_{#1}}
\newcommand{\itrsd}[1]{\alpha_{#1}}
\newcommand{\itail}{\Delta}

\newcommand{\ans}{w}
\newcommand{\dsize}{v}
\newcommand{\pvp}[1]{\prob_{\lowt}({#1})}
\newcommand{\pb}[1]{\prob({#1})}
\newcommand{\pevpp}{\prob_{e}^{\text{c} \rightarrow \text{p}}}
\newcommand{\pepvp}{\prob_{e}^{\text{p} \rightarrow \text{c}}}
\newcommand{\pemvp}{\prob_{e}^{\text{m} \rightarrow \text{c}}}
\newcommand{\pevpm}{\prob_{e}^{\text{c} \rightarrow \text{m}}}
\newcommand{\pis}[1]{\prob^{(0)}_{#1}}
\newcommand{\pd}[1]{\prob^{(\person)}_{#1 \mid \text{d}}}
\newcommand{\pnd}[1]{\prob^{(\person)}_{#1 \mid \text{nd}}}

\IEEEoverridecommandlockouts

\begin{document}

\title{Near-Optimal Stochastic Threshold Group Testing}
\author{\IEEEauthorblockN{C. L. Chan$^\ast$, S. Cai$^\ast$, M. Bakshi$^\ast$, S. Jaggi$^\ast$, and V. Saligrama$^+$\\
The Chinese University of Hong Kong$^\ast$, Boston University$^+$}\\
}

\maketitle

\begin{abstract}
We formulate and analyze a stochastic threshold group testing problem motivated by biological applications. Here
a set of $n$ items contains a subset of $d \ll n$ defective items. Subsets (pools) of the $n$ items are tested -- the test outcomes are negative, positive, or stochastic (negative or positive with certain probabilities that might depend on the number of defectives being tested in the pool), depending on whether the number of defective items in the pool being tested are fewer than the {\it lower threshold} $l$, greater than the {\it upper threshold} $u$, or in between. The goal of a
 {\it stochastic threshold group testing} scheme is to identify the set of $d$ defective items via a ``small'' number of such tests. In the regime that $l = o({d})$ we present schemes that are computationally feasible to design and implement, and require near-optimal number of tests (significantly improving on existing schemes).
Our schemes are robust to a variety of models for probabilistic threshold group testing.
\end{abstract}

\IEEEpeerreviewmaketitle

\section{Introduction}

\noindent \textbf{Classical Group Testing:} The set ${\cal N}$
of $n$ items contains a set ${\cal D}$ of $d$ ``defectives'' -- here $d$ is assumed to be $o(n)$.
The classical version of the group-testing problem was first considered
by Dorfman in 1943~\cite{Dorfman:1943}
as a means of identifying a small number of diseased individuals from
a large population via as few {}``pooled tests''
as possible. In this scenario, blood from a subset of individuals
is pooled together and tested -- if none of the individuals being tested in a pool have the disease the
test outcome is {}``negative'', else it is ``positive''.
In the non-adaptive group testing problem, each test is designed independently
of the outcome of any other test, whereas for adaptive group-testing problems, the testing procedure may be conducted sequentially. For both problems, ${\cal O}(d\log(n))$ tests are known to be necessary and sufficient -- a good survey of some of the algorithms
and bounds can be found in the books by Du and Hwang~\cite{DuH:00, NGT}
and the paper by Chen and Hwang~\cite{ChenH:2008}.

\noindent \textbf{Threshold Group Testing: } In this work we focus on a generalization of the classical group testing problem called {\it threshold group testing}, first considered by Damaschke~\cite{Damaschke:2006}.
The difference is that the outcome of each pooled test is {}``positive''
if the number of defectives in the test is no smaller than the {upper threshold}
(denoted $u$), is {}``negative'' if no larger than the { lower threshold}
(denoted $l$) defectives were contained in the test, and otherwise it
is arbitrary (``worst-case''). Clearly, when $u=1$ and $l=0$, this reduces to the classical
group testing problem. There are other generalizations of classical group testing \cite{ChiLY:2011,CheB:2011,EmaM:2012}.
Applications of the
threshold group testing model include the problem of reconstructing a hidden
hypergraph \cite{Cheraghchi:2010,
ChenDH:2007, ChangCFS:2011,ChenF:2009}, and a searching problem called
``guessing secrets''~\cite{Damaschke:2006, AloGKS:2007}.

%
%
%




The first adaptive algorithm for threshold group testing was proposed in~\cite{Damaschke:2006}. {When the {\it gap} $g$ (defined as $u-l-1$, the difference between the upper and lower thresholds) equals $0$,} the number of tests in~\cite{Damaschke:2006}
for identification of the set of defectives is ${\cal O}\left(\left(d+u^{2}\right)\log n\right)$.
When the gap $g\neq0$, the number of tests required by~\cite{Damaschke:2006} scales as ${\cal O}(dn^{b}+d^{u})$,
if $g+(u-1)/b$ misclassifications are allowed (here $b>0$
is an arbitrary constant), with polynomial-time decoding complexity.
The work of~\cite{ChenF:2009} showed that ${\cal O}\left(ed^{u+1}\log(n/d)\right)$ non-adaptive threshold tests suffice to identify the set of defectives with up to $g$ misclassifications
and $e$ erroneous tests allowed. The computational complexity of decoding is ${\cal O}\left(n^{u}\log n\right)$ for fixed $(d,e)$.
In \cite{Cheraghchi:2010}, instead of the strongly disjunct matrices used in~\cite{ChenF:2009}, a probabilistic construction of a weaker version of disjunct matrices is used to reduce the number
of tests from ${\cal O}\left(d^{u+1}\log(n/d)\right)$ to ${\cal O}\left(d^{g+2}\log d\log(n/d)\right)$.
{Also, two explicit constructions with number of tests equaling
${\cal O}\left(d^{g+3}\left(\log d\right)\mbox{quasipoly}\left(\log n\right)\right)$
and ${\cal O}\left(d^{g+3+\beta}\mbox{poly}\left(\log n\right)\right)$ (for arbitrary $\beta>0$) are proposed.} However, the computational complexity of decoding is not addressed.
Also,~\cite{AhlDL:2011} draws a connection between ``threshold codes'', non-adaptive threshold group testing, and a model called ``majority group testing''.


\noindent
{\bf (A) Worst-case Model:} If the number of defective items in a pool is between the upper and lower thresholds (``in the gap''), then the test outcome is assumed to be {\it arbitrary}. Algorithms must therefore be designed to account for a malicious adversary that can set test outcomes to maximally confuse the threshold group testing scheme.

\noindent
{\bf (B) Zero-error (with misclassifications):} The algorithm is required to guarantee (with probability $1$), that the output is ``correct'' (it contains the set of defective items, up to a certain number of misclassifications).
(A fundamental consequence of these two models assumptions is that if the gap $g=u-l-1>0$, the set of defectives {\it cannot} be exactly identified -- {\it regardless} of what algorithm is used, one can only reconstruct the set of defective items up to a certain number of misclassifications~{\cite{Damaschke:2006}}).

\noindent \textbf{Stochastic Threshold Group Testing: } We relax these aspects of the conventional setting. In particular, we relax the worst-case model to a stochastic model. We seek probabilistic guarantees instead of the absolute zero error guarantees. 

\begin{figure}
\begin{minipage}[t]{0.48\linewidth}
\centering
\includegraphics[width=1\linewidth]{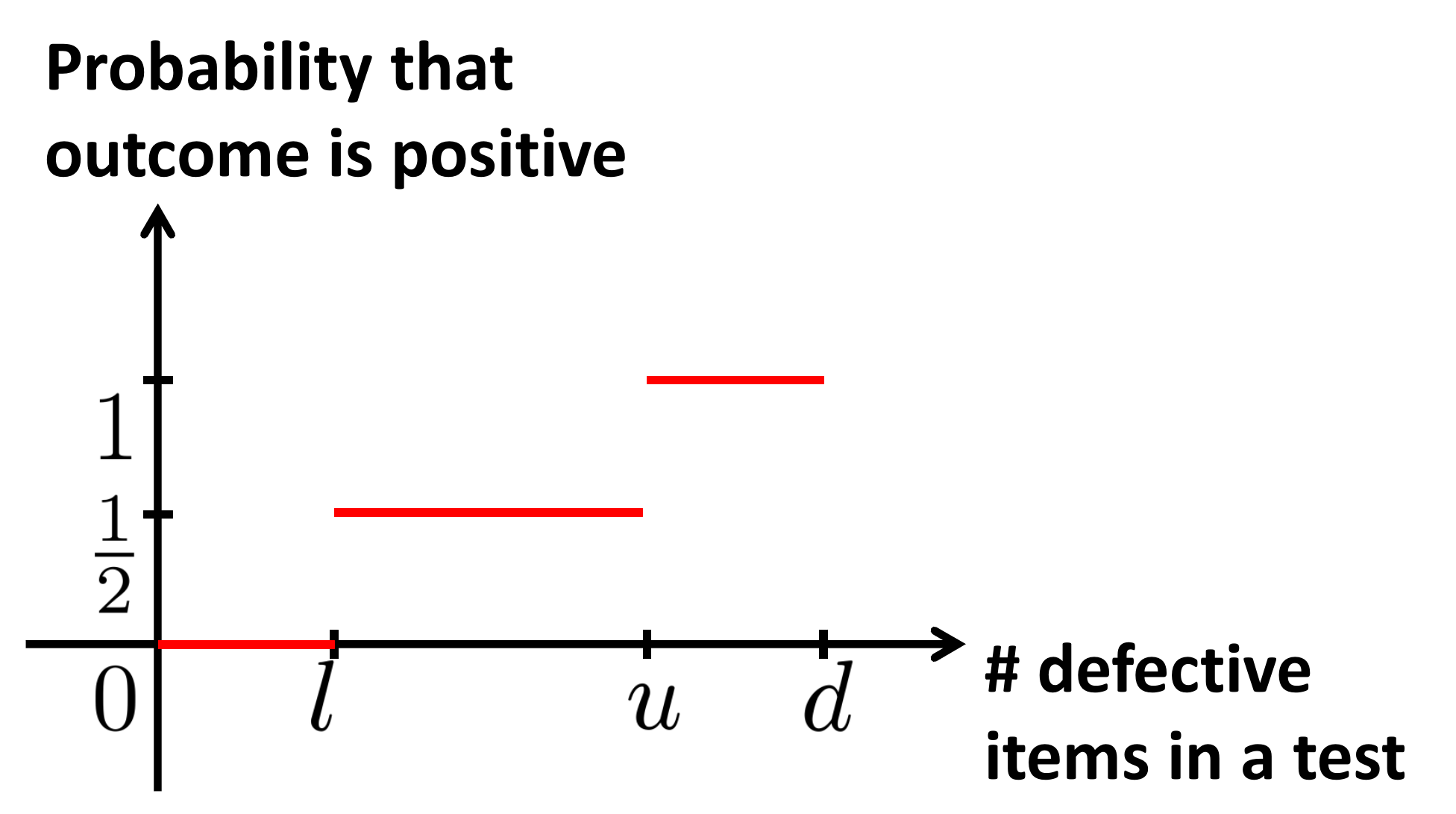}
\caption{\textbf{Bernoulli gap stochasticity}: If the number of defectives present in a test is between the thresholds, the probability that the outcome is positive equals $1/2$.}
\label{fig:model1}
\end{minipage}%
\hspace{0.1in}
\begin{minipage}[t]{0.48\linewidth}
\centering
\includegraphics[width=1\linewidth]{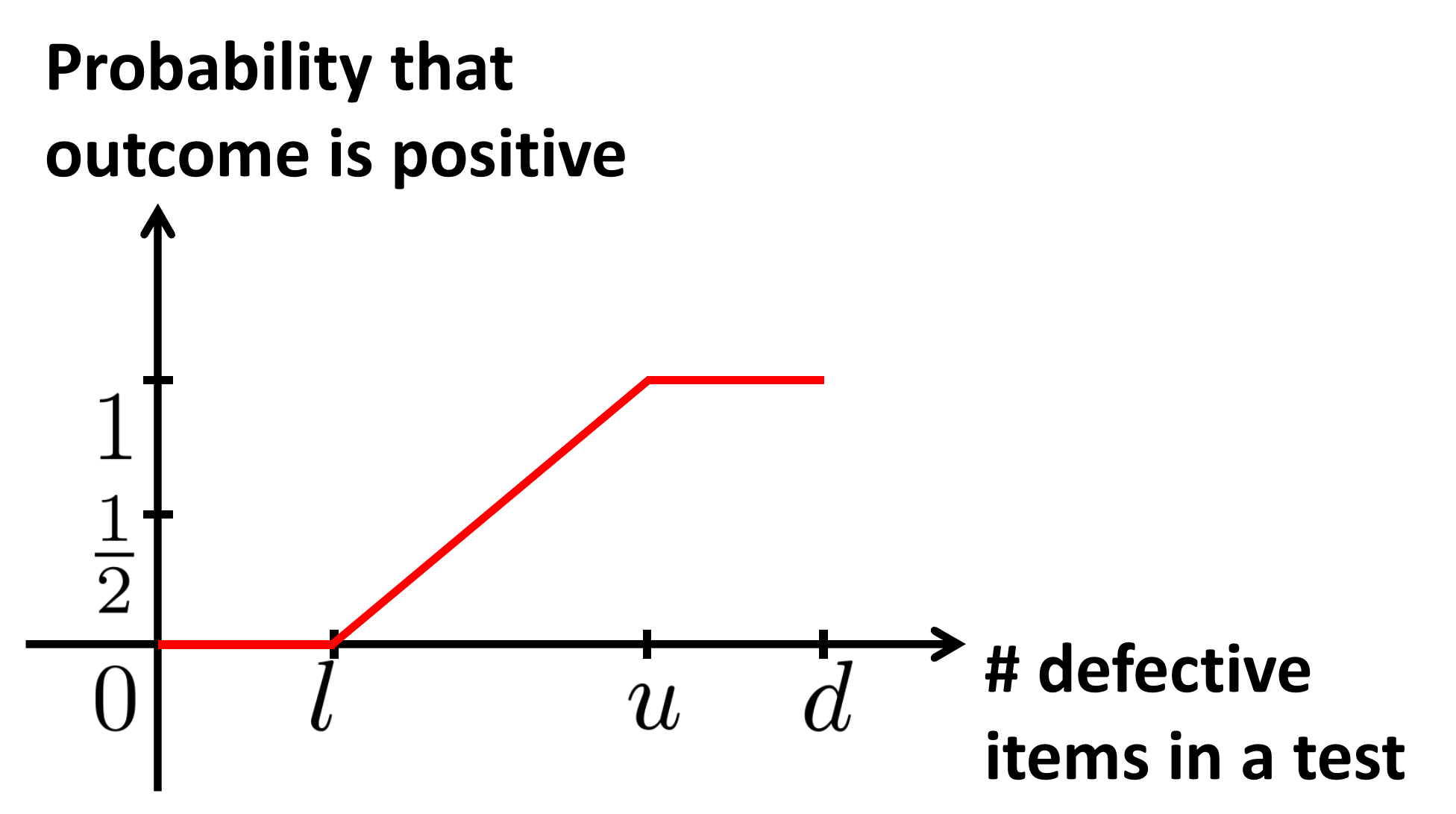}
\caption{\textbf{Linear gap stochasticity}: If the number of defectives present in a test is between the thresholds, the probability that the outcome is positive increases linearly.}
\label{fig:model2}
\end{minipage}
\end{figure}

\noindent 
{\bf (A) Stochastic Model:}  This setup is motivated by a class of biological applications~\cite{Maltezos} where the test outcomes are observed to be random whenever the number of defectives in a pool falls within a given range. We consider two models. For the first model, we assume that the outcome of a test is equally likely to be positive or negative whenever the number of defectives in a pool is in the range $(l,u)$.  In our second model, the probability of a test outcome being positive depends on the number of defective items in the test, and for concreteness, we assume that this dependence scales linearly from $l$ to $u$ (though our results hold for more general models as well\footnote{In fact, as long as there is a statistical difference between the probability of a positive test outcome when the number of defective items is within the range $(l,u)$, and outside this range, our approach works. Due to space limitations, in this work we focus on the two models in Figures~\ref{fig:model1} and~\ref{fig:model2}.}). These two models are represented in Figures~\ref{fig:model1} and~\ref{fig:model2}.

\noindent 
{\bf (B) Probabilistic Guarantee:} We allow for a ``small'' probability of error for our algorithm, where this probability is both with respect to the randomness of the measurements within the gap, and the test design.

These ``natural'' information-theoretic relaxations in the model result in schemes that have significantly improved performance, compared to prior work. In particular, our schemes require far fewer tests than prior algorithms, and also admit computationally efficient decoding schemes. They also directly lend themselves to scenarios with zero gaps, and also to other models similar to group-testing, such as the Semi-Quantitative Group Testing~\cite{EmaM:2012}.

For the stochastic threshold group-testing problem we present three algorithms (\algoa, \algob, and \algoc, respectively for the non-adaptive problem with Bernoulli gap stochasticity, adaptive  problem with Bernoulli gap stochasticity, and non-adaptive  problem with linear gap stochasticity).
Our results are summarized as follows.


\begin{theorem}{\bf (Non-adaptive algorithm with Bernoulli gap model)}
For $\lowt=o(\dfct)$, {\algoa} with error probability at most $\error{}$ requires $(4e^8\ln(2)/\pi^2)\ln(1/\error{})\sqrt{\lowt}\dfct\ln(\bl) + \bigo{\ln(1/\error{})\dfct\sqrt{\lowt}}$ tests and computational complexity of decoding $\bigo{\bl \ln(\bl) + \bl\ln(1/\error{})}$.
\label{thm:tgt1}
\end{theorem}

\begin{theorem}{\bf (Two-stage Adaptive algorithm)}
For $\lowt=o(\dfct)$, {\algob} with error probability at most $\error{}$ requires $16e^2\dfct\ln(\bl) + \bigo{\ln(1/\error{})\dfct}$ tests and computational complexity of decoding $\bigo{\bl\ln(\bl) + \bl\ln(1/\error{})}$.
\label{thm:tgt2}
\end{theorem}

\begin{theorem}{\bf (Non-adaptive algorithm with linear gap model)}
{\algoc} with error probability at most $\error{}$ requires $\bigo{\gap^2\dfct\ln(\bl)} + \bigo{\ln(1/\error{})\dfct}$ tests and computational complexity of decoding $\bigo{\gap^2 \bl \ln(\bl) + \bl\ln(1/\error{})}$.
\label{thm:tgt3}
\end{theorem}

{\bf Remark:} Note that the number of tests required by our algorithms are, in general, much smaller than those required by prior works -- this demonstrates the power of using the stochasticity that may naturally be inherent in the measurement model.

\section{Intuition}

\begin{table}
\centering
\begin{tabular*}{0.9\linewidth}{c | l }
\multicolumn{2}{c}{Model parameters} \\
\hline
\hline
${\cN}$ & The set of all items. \\
$\bl$ & The total number of items, $\bl = |\cN|$. \\
${\cD}$ & The unknown subset of defective items. \\
$\dfct$ & The total number of defective items, $\dfct = |\cD|$. \\
$\inp_\person$ & The binary indicator variable corresponding to the $\person$-th item. \\
$\lowt$ & The lower threshold. \\
$\upt$ & The upper threshold. \\
$\gap$ & The gap $(\gap\triangleq \upt - \lowt - 1)$ between the two thresholds.
 $\lowt$ and $\upt$. \\
$\test$ & The total number of tests. \\
\hline
\multicolumn{2}{c}{} \\
\multicolumn{2}{c}{Algorithmic parameters} \\
\hline
\hline
$\parta_\dstr$ & The $\dstr$-th division of $\cN$ into separate regions.\\
$\bin_\dstr$ & The complement of $\parta_\dstr$ that contains \reference groups. \\
$\bnum$ & The total number of divisions of $\parta$. \\
$\rfrn_{\dstr, \family}$ & The $\family$-th \reference group in $\bin_\dstr$. \\
$\rrpt$ & The total number of \reference groups in each $\bin_\dstr$. \\
$\cidct_\stage$ & The set of \indicator groups in the $\stage$-th family/partition of $\cN$ \\
$\idct_{\stage, \pool}$ & The $\pool$-th \indicator group in the $\stage$-th family. \\
$\irpt$ & The total number of families for \indicator groups. \\
$\idct^{(0)}_\stage$ & A randomly picked \indicator group from $\cidct_\stage$. \\
$\cidct^{(\person)}$ & The subset of \indicator groups from $\bigcup_\stage \cidct_\stage$ that includes $\inp_\person$. \\
$\outptu{\person}{\dstr, \family}{\stage}$ & The test outcome when the group $\rfrn_{\dstr, \family} \cup \idct^{(\person)}_{\stage}$ is measured. \\
\hline
\end{tabular*}
\vspace{0.1in}
\caption{Notation used frequently in this paper. We use calligraphic notation to denote sets, and boldface calligraphic notation to denote sets of sets.}
\label{table:notation}
\end{table}

\begin{figure}[htbp]
   \centering
   \includegraphics[width=6cm]{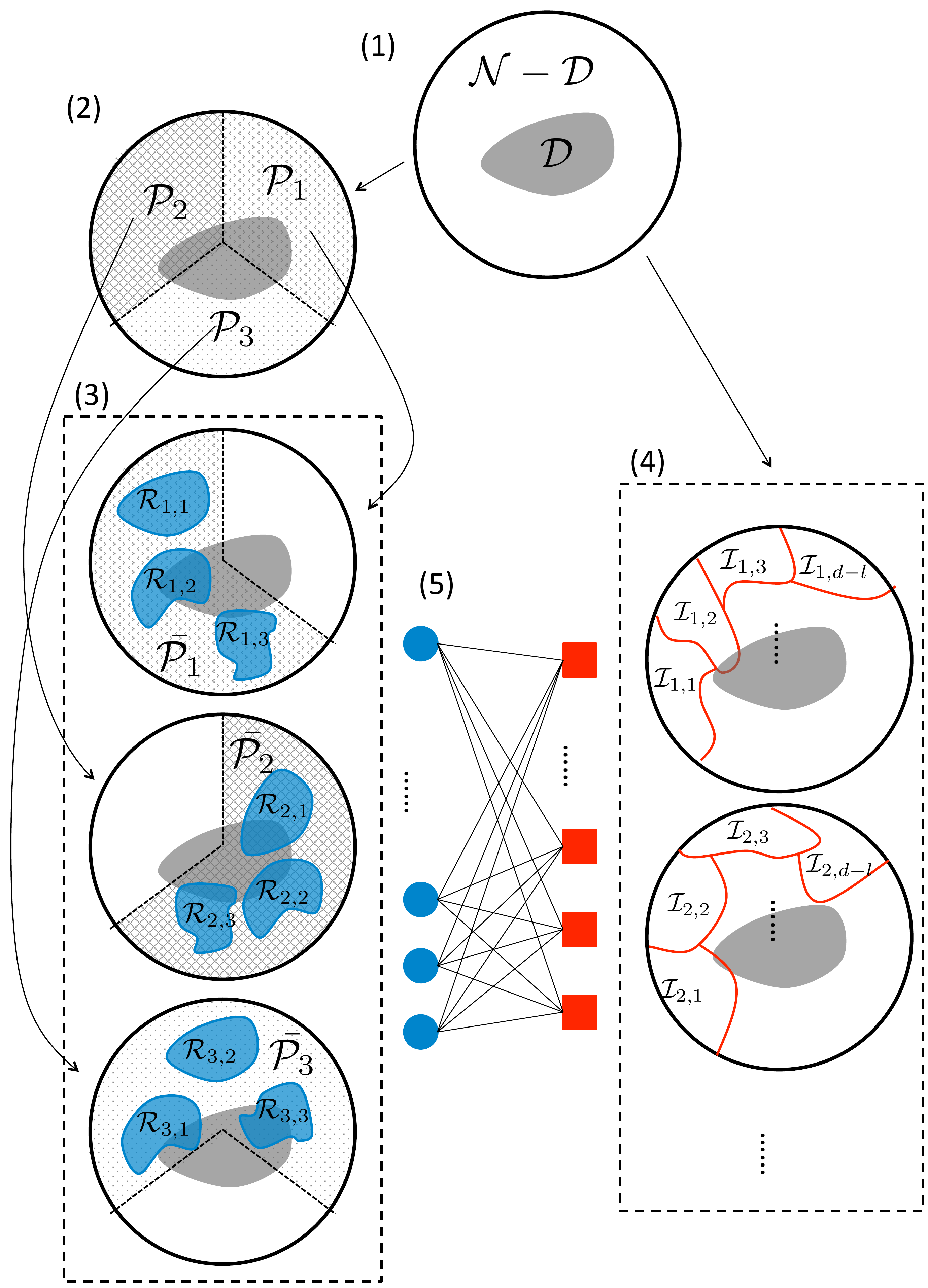} 
   \caption{ An example illustrating \algoa's testing/encoding scheme: (1) The population $\cN$ of $n$ items contains a small subset $\cD$ of $d$ defective items (the shaded region). (2) $\cN$ is partitioned equally into $\parta_1$, $\parta_2$ and $\parta_3$ (in this case $\bnum = 3$). (3) From each $\bin_\dstr$ (the complement of $\parta_{\dstr}$), $\rrpt$ \reference groups of size $nl/d$ (in this example, $\rrpt$ also equals $3$) are picked uniformly at random. (4) Independently of all prior choices, $\irpt$ families of indicator groups $\cidct_i$ are chosen. In each family, $\cN$ is partitioned uniformly at random into \indicator groups of size $\iratio \bl/(\dfct-\lowt)$ each (here $\iratio$ is a code-design parameter whose value is specified later). (5) The complete bipartite graph shows the ``cross-product'' design of threshold tests of the form $\crfrn \times \cidct$, in which left ({blue}) nodes represent \reference groups, right ({red}) nodes represent \indicator groups, and edges represent threshold tests corresponding to the {union of} items denoted by the connected nodes (one reference group with one indicator group). }
   \label{fig:encode}
\end{figure}

\begin{figure}[htbp]
   \centering
   \includegraphics[width=6cm]{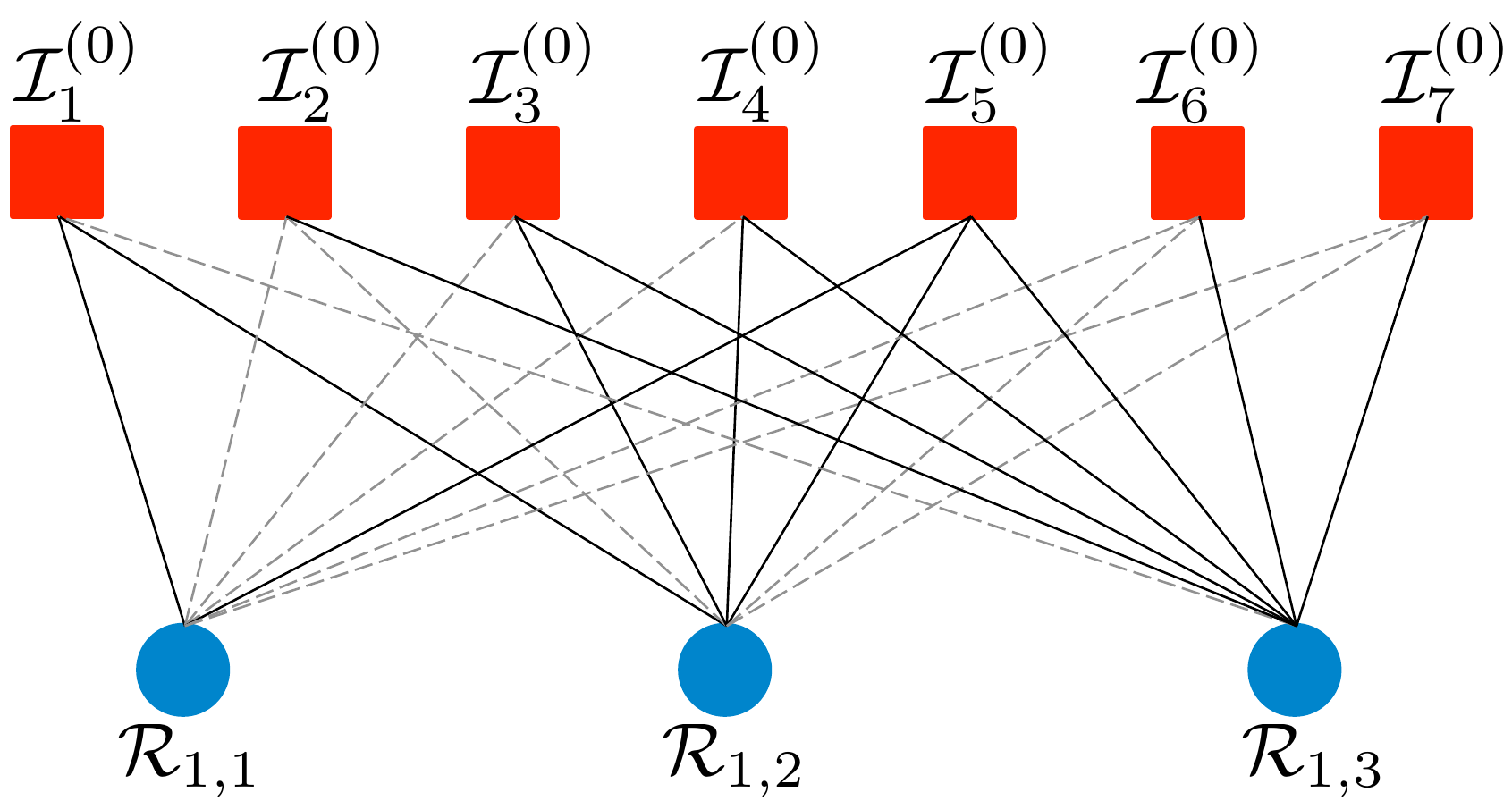} 
   \caption{ An example illustrating \algoa's scheme to estimate whether a particular reference group is a \critical~\reference group or not. The top nodes are a set of randomly picked \indicator groups from different families. The bottom nodes represent \reference groups picked uniformly at random in a particular division (say $\bin_1$). Each bottom node connects to all the top nodes by an edge {whose type indicates} the test outcome when the {union of} items denoted by its connected nodes are measured. A solid edge indicates a positive outcome; conversely, a dashed edge indicates a negative outcome. In this example, $\rfrn_{1,1}$ has less than $\lowt$ defective items, and is unlikely to hit too many \indicator groups with enough defective items such that their union has more than $\lowt$ defective items (in this example, only $2$ out of $7$ test outcomes are positive). At the other extreme, $\rfrn_{1,3}$ has more than $\lowt$ defective items, and has a ``fairly high'' probability to give a positive outcome (in this example, $6$ out of $7$ test outcomes are positive). In between these two extremes, it is expected that a ``{\critical}'' \reference group (with exactly $\lowt$ defective items) would produce positive test outcomes with an intermediate ratio (in this example, for $\rfrn_{1,3}$, $4$ out of $7$ test outcomes are positive, and say the expected empirical number of positive test outcomes for a critical \reference group is in the range $3$ to $5$). Hence, the decoder declares that (only) $\rfrn_{1,2}$ is critical.}
   \label{fig:decode1}
\end{figure}

\begin{figure}[htbp]
   \centering
   \includegraphics[width=6cm]{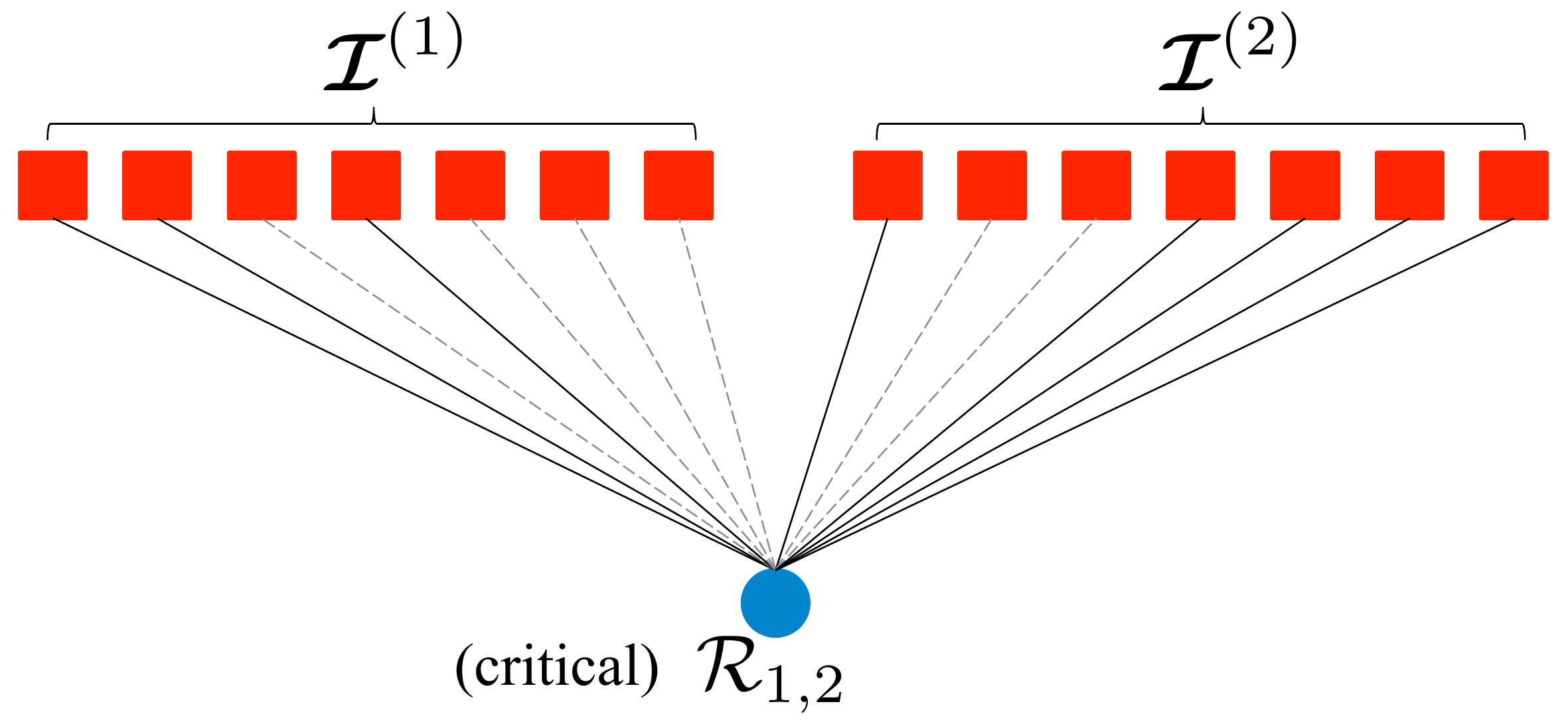} 
   \caption{ An example illustrating \algoa's decoding scheme to determine whether particular items are defective or not. The sets of top nodes denote the sets of \indicator groups that include a particular item $\inp_\person$ -- in this example, we consider only the indicator groups corresponding to $x_1$ and $x_2$. The bottom node represent a \critical~\reference group. An edge has the same definition as in Figure~\ref{fig:decode1}, {\it i.e.}, a threshold test corresponding to the union of the items in the \indicator group and the \critical~\reference group. A test comprising of an indicator group corresponding to a defective item $\inp_\person$ with a \critical~\reference group will {\it always} have more than $\lowt$ defective items. Hence such tests are expected to give a higher ratio of positive test outcomes, than if the item $\inp_\person$ is non-defective (since at least some of the corresponding indicator drops will not contain any defective items -- if one chooses the side of the indicator groups carefully (about ${\cal O}(n/d)$), this will happen in a constant fraction of indicator groups. In this example, the decoder declares the 1st item to be non-defective (since ``few'' of the test outcomes are positive), and the 2nd item to be defective (since ``many'' of the test outcomes are positive). }
   \label{fig:decode2}
\end{figure}

To build intuition into our proof techniques consider the Bernoulli Stochastic Model described in Sec.~1 and Fig.~\ref{fig:model1}. We note the discrete transition in terms of \emph{distribution} of test outcomes for pools consisting of $l+1$ defectives relative to those that contain $l$ defectives. If negative outcomes are labeled zero and positive outcomes labeled one, the test outcomes are identically zero for pools containing exactly $l$ defectives. So the distribution of test outcomes is concentrated at zero. For pools containing $l+1$ defectives the distribution of test outcomes is split equally at zero and one. We can exploit this aspect of the model in the following way. Suppose we had a pool, $\rfrn^*$ consisting of exactly $l$ defectives then one could test whether or not an item, $x_j \not \in  \rfrn^*$ is defective by augmenting ${\cal R}^*$ with $x_j$ and testing the new pool ${\cal R}^* \bigcup \{x_j\}$. 

To exploit this idea we have to account for several issues. First, we do not really have a candidate pool ${\cal R}^*$. Second, with this naive strategy, the number of tests would grow with the number of items even when we have a candidate pool ${\cal R}^*$ consisting of $l$ defectives. 

To address these requirements we construct two distinctive collections of pools based on random designs. The reference group collection, $\crfrn$,  is a collection of $R$ pools, each with $nl/d$ items, such that at least one among the $R$ pools has exactly $l$ defective items in it. The idea is that with high probability one among the $R$ pools contains the critical candidate $\rfrn^*$. The second collection, the transversal design, is a family $\cidct$ of sub-collections of size $I$. Each sub-collection within this family consists of $O(d)$ disjoint pools indexed as $\idct_{i,k}$. Each disjoint pool within a sub-collection is referred to as an indicator group. Consequently, each item appears only once within any sub-collection and $I$ times within the entire family. The indicator group collection serves the role of an item $\{x_j\}$ described in the preceding paragraphs. 

Our algorithm is based on augmenting each indicator group within a sub-collection with the reference group collections to form pools that are then tested. 
The idea of transversal design is not new and has been used before in conventional group testing~\cite{BalBTK:1996} as well. The novelty here is the cross-product, namely, testing indicator groups against a reference group collection resulting in $\crfrn \times \cidct$ pools. The question arises as to how to construct these collections and how to find defectives given the test outcomes. 

To construct $\crfrn$ we begin by noting that if one chooses a {\it random} group of size $nl/d$, with probability about $1/\sqrt{l}$ it has exactly $l$ defective items in it. This is because of the fact that the expected number of defective items in such a group is exactly $l$, and standard analysis using Stirling's approximation of the hypergeometric distribution corresponding to the number of defective items in a group of size $nl/d$ implies that the probability of hitting this expectation scales as $1/\sqrt{l}$. This means that if one chooses about $\rrpt = {\cal O}(\sqrt{l})$ ``candidate reference groups'' $\rfrn_\family$, each of size $nl/d$, then ``with high probability'' at least one group $\rfrn^*$ will be ``critical'' (have exactly $l$ defective items in it).
%
To summarize, we select ${\cal O}(\sqrt{l})$ candidate reference groups of size $nl/d$ each, and ${\cal O}(d\log(n))$ indicator groups of size about $n/d$ each. We then perform threshold group-tests on every pair of the form $\rfrn_\family \cup \idct_{}$, for a total of about ${\cal O}(\sqrt{l}d\log(n))$ (non-adaptive) tests.

Our decoding algorithm hinges on identifying the critical candidate(s) $\rfrn^*$. 
To do so we make use of the statistical difference between reference groups that are critical, and those that are not. When $\rfrn$ and a randomly-picked \indicator group are tested together, the probability of observing a positive outcome is an increasing function of the number of defective items in $\rfrn$. Hence the decoder performs \emph{matching and quantization}, as follows.
%
%
 For a large set of randomly-picked \indicator groups $\cidct$, each \indicator group is tested with $\rfrn$. The decoder computes the empirical fraction of tests with positive outcomes, and estimates whether $\rfrn$ is critical or not by comparing this empirical fraction with a pre-computed expected fraction for a \critical~\reference group. By the Chernoff bound, one can concentrate the variation of the empirical concentration around the expectation quite tightly, and hence, with high probability, estimate whether or not a given reference group is critical.

Another challenge remains -- namely, how does one identify the defective items {\it within} $\rfrn^*$?
The solution proposed is to divide $\cN$ into $\bnum$ disjoint divisions $\parta_\dstr$, and sample \reference groups from every $\bin_\dstr$, the complement of $\parta_\dstr$. If the decoder discovers a \critical~\reference group $\rfrn^*_\dstr$ from $\bin_\dstr$, all items in $\cN\setminus\rfrn^*_\dstr$ ($\supseteq \parta_\dstr$) are decodable. Furthermore, if the decoder discovers a \critical~\reference group $\rfrn^*_\dstr$ from every $\bin_\dstr$, all items in $\cN$ (including those in any \critical~\reference group $\rfrn^*_\dstr$) are decodable since $\cN=\bigcup \parta_\dstr \subseteq \bigcup (\cN \setminus \rfrn^*_\dstr) \subseteq \cN$.

If one is allowed to perform {\it adaptive} threshold group tests (even just two stages), then one can significantly reduce the number of tests required. The idea is to use the first stage to identify critical reference groups (have exactly $l$ defectives) and then use {\it only} those reference groups in subsequent stages. Hence one reduces the overall number of tests required by a factor of $\sqrt{l}$, since one no longer needs to test all cross-products between all reference groups and all indicator groups.

In the case that the probability of positive test outcomes scales linearly in the gap (as in Figure~\ref{fig:model2}), we note two  conflicting factors at play. On the one hand, suppose a test contains $v$ defective items, now there is a large range  of values ($v$ can be all the way from $l$ to $u-1$) for which the probability of positive test outcomes differs between tests with $v$ defectives and tests with $v+1$ defectives. Hence any such group can be used as a proxy for a critically thresholded group (instead of demanding that such critical groups contain {\it exactly} $\l$ defective items). If the gap $g$ is ``reasonably large'', then in fact choosing a reference group of size $n(u+l)/(2d)$ results in a group falling within this range with high probability. Hence one does not need ${\cal O}(\sqrt{l})$ reference groups to find good ones -- a constant number suffice.
On the other hand, the statistical difference in the empirical probability of observing positive test outcomes now only changes very slightly, if a group contains $v$ defective items, and if it contains $v+1$ defective items. This difference in fact scales as $1/g$. To be able to reliably detect such a slight change, one has to perform about a factor $g^2$ more tests. Hence in this linear gap model of stochastic threshold group testing, the number of tests required differs from the Brenoulli gap model by a factor of $g^2/\sqrt{l}$.

\section{Algorithm for Theorem~\ref{thm:tgt1}}
We now formally describe the \algoa~algorithm that meets the conditions of Theorem~\ref{thm:tgt1}. 

\noindent {\bf Encoder/Testing scheme:}
\begin{enumerate}
\item{
Let $\bratio=(\dfct+\lowt)/(2\dfct)$ and $\bnum=1/(1-\bratio)$.
Partition $\cN$ into {$\bnum$} disjoint sets $\{ \parta_\dstr: \dstr = 1, 2, \cdots, \bnum \}$, each of equal size $\bl / \bnum$. Denote the complement of $\parta_\dstr$ by $\bin_\dstr$. Note that every $\bin_\dstr$ is of size $\bratio\bl$. The encoder generates {\em $\family$-th \reference group} $\rfrn_{\dstr, \family}$ by randomly picking $\bl\lowt/\dfct$ distinct items from $\bin_\dstr$. This process is repeated $\rrpt$ times for every $\bin_\dstr$ so that the encoder obtains a set of \reference groups $\crfrn= \{\rfrn_{\dstr, \family} : \dstr =1,2, \cdots, \bnum ; \family = 1, 2, \cdots, \rrpt \}$.
}
\item{
Let $\iratio \in (0,1]$. For each family $\stage \in \{1, 2, \cdots, \irpt\}$, the encoder generates a random partition $\{ \cidct_\stage\} = \{ \idct_{\stage, \pool}: \pool = 1,2, \cdots, \dfct - \lowt \}$ of $\cN$, where each $\idct_{\stage, \pool}$ is of size $\iratio\bl/(\dfct-\lowt)$, and we call it an {\em \indicator group}. $\cidct= \bigcup \cidct_{\stage}$ represents the set of \indicator groups.
}

\item{
For every pair of $\rfrn_{\dstr, \family}$ and $\idct_{\stage, \pool}$, the encoder performs a threshold test on $\rfrn_{\dstr, \family} \cup \idct_{\stage, \pool}$.
}
\end{enumerate}

\noindent {\bf Decoder:}
\begin{enumerate}
\item{
For each $\stage = 1, 2, \cdots, \irpt$, let $\idct^{(0)}_\stage$ be a randomly picked indicator group from $\{ \cidct_\stage \}$, and let $\cidct^{(0)} = \{ \idct^{(0)}_\stage : \stage = 1, 2, \cdots, \irpt \}$. Let $\outptu{0}{\dstr, \family}{\stage}$ be the test outcome when the group $\rfrn_{\dstr, \family} \cup \idct^{(0)}_{\stage}$ is measured, and $\rtrsd{\dsize} = \prob \left ( \outptu{0}{\dstr, \family}{\stage} = 1 \bigm | |\rfrn_{\dstr, \family} \cap \cD| = \dsize \right )$. The decoder declares $\rfrn_{\dstr, \family}$ to be \emph{\critical} if
\begin{align*}
|\cidct^{(0)}|\rtrsd{\lowt}(1-\rtail{{<\lowt}}) \leq \sum_\stage{\outptu{0}{\dstr, \family}{\stage}} \leq |\cidct^{(0)}|\rtrsd{\lowt}(1+\rtail{{>\lowt}})
\end{align*}
where $\rtail{{<\lowt}} = (\rtrsd{\lowt} - \rtrsd{{\lowt-1}})/(2\rtrsd{\lowt})$ and $\rtail{{>\lowt}} = (\rtrsd{{\lowt+1}} - \rtrsd{\lowt})/(2\rtrsd{\lowt})$. 
That is, the decoder declares a \reference group to be {\critical} if the empirically observed fraction of positive test outcomes involving that \reference group ``is close to'' the ``expected value'' ($\rtrsd{\lowt}$). The probability of this event is calculated in Lemma~\ref{thm:decvp}.
}
\item{
For each $\stage = 1, 2, \cdots, \irpt$, let $\idct^{(\person)}_\stage$ be a indicator group from $\{ \cidct_\stage \}$ that includes $\inp_\person$, and let $\cidct^{(\person)} = \{ \idct^{(\person)}_\stage : \stage = 1, 2, \cdots, \irpt \}$. Let $\outptu{\person}{\dstr, \family}{\stage}$ be the test outcome when the group $\rfrn_{\dstr, \family} \cup \idct^{(\person)}_{\stage}$ is measured, and $\itrsd{\ans} = \prob \left ( \outptu{\person}{\dstr, \family}{\stage} = 1 \bigm | |\rfrn_{\dstr, \family} \cap \cD| = \lowt, \inp_{\person} = \ans \right )$. Note that $\inp_\person$ is a binary indicator variable taking values $1$ or $0$ depending on whether the item is defective or not. 
For every \critical~$\rfrn_{\dstr, \family} \subseteq \bin_\dstr$ and $\inp_\person \in \parta_\dstr$, if $\sum_\stage{\outptu{\person}{\dstr, \family}{\stage}} \leq |\cidct^{(\person)}|\itrsd{0}(1+\itail)$ where $\itail = ( \itrsd{1} - \itrsd{0} ) / (2\itrsd{0})$, the decoder declares $\inp_j$ to be non-defective, else declares it to be defective. That is, the decoder declares an item to be non-defective if the empirically observed fraction of positive test outcomes involving that item ``is close to'' the ``expected value'' ($\itrsd{0}$). The probability of this event is calculated in Lemma~\ref{thm:decinp}.
}
\end{enumerate}

\section{Proof {of Theorem \ref{thm:tgt1}}}

\begin{definition}
Hypergeometric distribution describes the probability of picking $\dsize$ defective items when we pick $s$ distinct items from $n$ items with $d$ defectives, The probability mass function is given by
\begin{equation*}
\prob(\dsize,s,n,d) = \frac{ \binom{d}{\dsize} \binom{n-d}{s-\dsize} }{ \binom{n}{s} }
\label{def:hyg}
\end{equation*}

\end{definition}

\begin{lemma}
The probability of picking $\lowt$ defective items when we sample $(\bl \lowt)/\dfct$ items from $\bl$ items with $\dfct$ defective items is $\Omega \left( \frac{1}{\sqrt{\lowt}} \sqrt{\frac{\dfct}{\dfct-\lowt}} \sqrt{\frac{\bl}{\bl-\dfct}} \right )$.
\label{thm:pvp}
\end{lemma}

\begin{figure}[htbp]
   \centering
   \includegraphics[width=8cm]{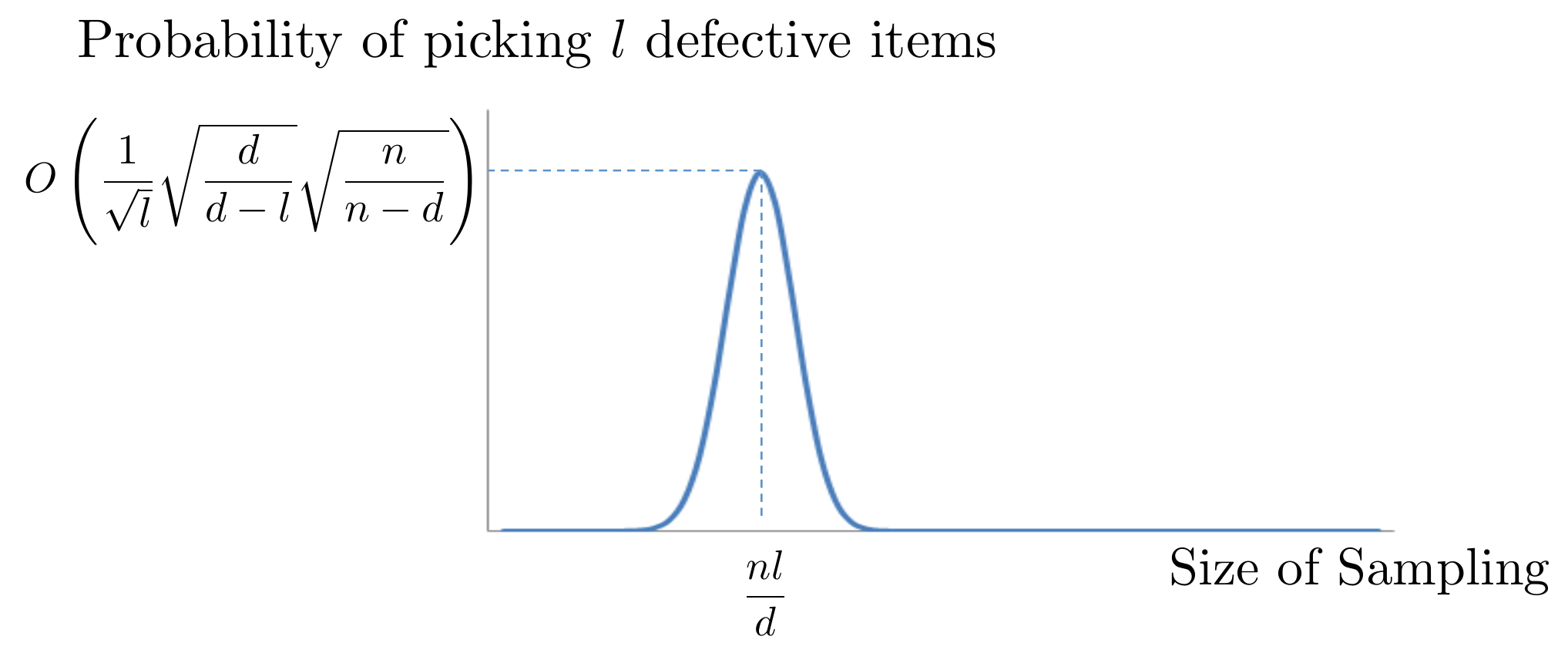} 
   \caption{Probability of sampling $\lowt$ defective items as a function of the size of the group undergoing a threshold test.}
   \label{fig:pmf}
\end{figure}

\begin{IEEEproof}
The probability that a test of a certain size has {\it exactly} $\lowt$ defective items scales according to the hypergeometric distribution given in Definition~\ref{def:hyg}. When the number of items in the test equals $\bl\lowt/\dfct$, this probability can be shown via Stirling's approximation~\cite{MitU:2005} that for all $n \in \mathbb{N}_+$, $1 \leq n!(2 \pi n)^{-1/2}(n/e)^{-n} \leq e(2\pi)^{-1/2}$, and therefore implying for all $k \in \mathbb{N}_+, k < n$, $ \sqrt{2\pi}/e^2 \leq \binom{n}{k} \sqrt{\frac{k(n-k)}{n}} \frac{k^k(n-k)^{n-k}}{n^n} \leq e/(2\pi)$ to scale as 
\begin{align}
\frac{\binom{\dfct}{\lowt}\binom{\bl - \dfct}{\frac{\bl\lowt}{\dfct}-\lowt}}{\binom{\bl}{\frac{\bl\lowt}{\dfct}}}
& \geq \frac{4\pi^2}{e^5}\frac{1}{\sqrt{\lowt}} \sqrt{\frac{\dfct}{\dfct-\lowt}} \sqrt{\frac{\bl}{\bl-\dfct}}. \label{eq:pvp00}
\end{align}
Note that the exponential terms from the Stirling's approximation of the binomial coefficients are exactly cancelled out in (\ref{eq:pvp00}). 
\end{IEEEproof}

\begin{lemma}
With probability at least $1-\error{2}$, for each $\dstr \in \{1,\cdots, \bnum\}$, every $\bin_\dstr$ has at least one {\critical} \reference group when
\begin{IEEEeqnarray*}{rCl}
\rrpt & > & \left (\ln \left ( \frac{1}{\error{2}} \right ) + \ln \left ( \frac{2\dfct}{\dfct-\lowt} \right ) \right ) \frac{e^6}{4\pi^2} \sqrt{\lowt} \sqrt{\frac{\dfct-\lowt}{\dfct+\lowt}} \sqrt{\frac{\bl-\dfct}{\bl}}
\end{IEEEeqnarray*}
\label{thm:allref}
\end{lemma}

\begin{IEEEproof}
Let $\event{1}(\dstr)$ be the event that a specific $\bin_\dstr$ has ``too many'' defective items, {\it i.e.}, $\left \| \bin_\dstr \cap \cD \right \| > (1\pm \shift{})\bratio\dfct$ for any particular $\dstr$. Let $\event{1}$ be the event corresponding to the union of $\event{1}(\dstr)$, {\it i.e.}, that at least one division has too many defective items. Let $\event{2}$ be the event that there exists a $\bin_\dstr$ which contains no critical \reference group. First, we compute the union bound of probability of $\event{1}$ for all $\dstr$ as
\begin{IEEEeqnarray}{rCl}
\bnum \prob \left ( \event{1} \right ) & < & \frac{4\dfct}{\dfct-\lowt}~\text{exp} \left ( - \frac {2(\bl+2)((\shift{}\bratio\dfct)^2 - 1)} {(\bl-\dfct+1)(\dfct+1)} \right ) \label{eq:pebc1} \\
& < & \frac{4\dfct}{\dfct-\lowt}~\text{exp} \left ( - (\shift{}\bratio)^2 \dfct \right ) \label{eq:pebc2}
\end{IEEEeqnarray}
Inequality (\ref{eq:pebc1}) follows from Hush's bound~\cite{HusS:2005}. Note that $\bratio=(\dfct+\lowt)/(2\dfct)$. When $\shift{}$ is greater than $\sqrt{((4\dfct)/(\dfct+\lowt)^2) \ln \left ( (4\dfct)/((\dfct-\lowt)\error{1}) \right )}=o(1/\sqrt{\dfct})$, (\ref{eq:pebc2}) is bounded from above by a constant $\error{1}$.

Let $\pvp{\dsize}=\prob(|\rfrn_{\dstr, \family} \cap \cD| = \lowt \bigm | |\bin_\dstr \cap \cD| = \dsize)$, {\it i.e.} the probability that we pick a {\critical} \reference from $\bin_\dstr$ given $\bin_\dstr$ contains $\dsize$ defective items. Let $\pb{\dsize}=\prob( |\bin_\dstr \cap \cD| = \dsize)$, {\it i.e.} the probability that we pick a {\critical} \reference from $\bin_\dstr$. We wish to compute $\pvp{(1 + \shift{}) \bratio \dfct}$ and $\pvp{(1 - \shift{}) \bratio \dfct}$.
The ratio of $\pvp{\bratio \dfct}$ to $\pvp{(1+\shift{}) \bratio \dfct}$ can be computed as
\begin{align}
 & \frac{\pvp{\bratio \dfct}}{\pvp{(1+\shift{}) \bratio \dfct}} \nonumber \\
= & \frac { \binom{\bratio \dfct}{\lowt} \binom{\bratio \bl-\bratio \dfct}{\frac{\bl\lowt}{\dfct}-\lowt} \bigg / \binom{\bl}{\frac{\bl\lowt}{\dfct}} } { \binom{(1+\shift{})\bratio \dfct}{\lowt} \binom{\bratio \bl-(1+\shift{})\bratio \dfct}{\frac{\bl\lowt}{\dfct}-\lowt} \bigg / \binom{\bl}{\frac{\bl\lowt}{\dfct}} } \label{eq:pvp0a} \\
= & \prod_{i=1}^{\bratio \dfct \shift{}} \left ( 1 - \frac{\lowt}{\bratio\dfct+i} \right ) \prod_{i=0}^{\bratio \dfct \shift{}-1} \left ( 1 + \frac{\frac{\lowt}{\dfct}(\bl-\dfct)}{(\bl-\dfct)(\bratio-\frac{\lowt}{\dfct})-i} \right ) \nonumber \\
 \leq & \left ( 1 - \frac{\lowt}{\bratio\dfct+\bratio \dfct \shift{}} \right ) ^{\bratio \dfct \shift{}} \left ( 1 + \frac{\frac{\lowt}{\dfct}(\bl-\dfct)}{(\bl-\dfct)(\bratio-\frac{\lowt}{\dfct})-\bratio \dfct \shift{}} \right )^{\bratio \dfct \shift{}} \nonumber \\
\leq &~\text{exp} \left ( {-\frac{\lowt\shift{}}{1+\shift{}}} \right ) \text{exp} \left ( { \frac{\lowt\shift{} \bratio(\bl-\dfct)}{(\bl-\dfct)(\bratio-\frac{\lowt}{\dfct})-\bratio \dfct \shift{}}} \right ) \label{eq:pvp3a}
\end{align}
Equality (\ref{eq:pvp0a}) follows by noting that $\pvp{\dsize}$ follows hypergeometric distribution. Inequality (\ref{eq:pvp3a}) follows from $(1+x)^y \leq e^{xy}$ for $|x|<1$ and $y \geq 0$. For $\shift{}=o(1/\sqrt{\dfct})$ and $\lowt=o(\dfct)$, (\ref{eq:pvp3a}) is bounded from above by $e$. The same technique applied to the ratio of $\pvp{\bratio \dfct}$ to $\pvp{(1- \shift{}) \bratio \dfct}$ implies that it is less than $e$. These bounds on these ratios, together with Lemma~\ref{thm:pvp} gives us that
\begin{align}
\pvp{(1 \pm \shift{}) \bratio \dfct} & = \frac{\pvp{(1 \pm \shift{}) \bratio \dfct}}{\pvp{\bratio \dfct}}\pvp{\bratio \dfct}  \nonumber \\
& \leq \frac{4\pi^2}{e^6}\frac{1}{\sqrt{\lowt}} \sqrt{\frac{\dfct+\lowt}{\dfct-\lowt}} \sqrt{\frac{\bl}{\bl-\dfct}} \label{eq:pvp5}
\end{align}

Finally, we bound the probability of $\event{2}$ (note that $\event{2}$ is defined in the beginning of the proof) occurring as
\begin{align}
\prob(\event{2}) & < \bnum \sum_{\dsize=0}^{\dfct} \pb{\dsize} \left ( 1 - \pvp{\dsize} \right )^{\rrpt} \nonumber \\
& < \bnum~\prob ( \event{1} ) + \bnum \sum_{\dsize=(1 - \shift{}) \bratio \dfct}^{(1 + \shift{}) \bratio \dfct} \pb{\dsize} \left ( 1 - \pvp{\dsize} \right )^{\rrpt} \nonumber \\
& < \bnum~\prob ( \event{1} ) + \bnum \left ( 1 - \pvp{(1 \pm \shift{}) \bratio \dfct} \right )^{\rrpt} \label{eq:allref2}
\end{align}
We now substitute equations (\ref{eq:pebc2}) and (\ref{eq:pvp5}) into (\ref{eq:allref2}). Note that for large enough $\dfct$, the constant $\error{1}$ (bounded from above by the quantity in (\ref{eq:pebc2}) can be made arbitrarily small. Hence, if we bound (\ref{eq:allref2}) from above by constant $\error{2}$, for large enough $\dfct$ and $\rrpt$, (\ref{eq:allref2}) can be made smaller than any $\error{2}$, and we obtain Lemma~\ref{thm:allref}.
\end{IEEEproof}

\begin{lemma}
With probability at least $1-\error{3}$, the decoder correctly determines whether a given $\rfrn_{\dstr, \family}$ is \critical~or not when
\begin{equation*}
\irpt > 8 e^2 \left ( \ln(\rrpt \bnum) + \ln \left ( \frac{1}{\error{3}} \right ) \right )
\end{equation*}
\label{thm:decvp}
\end{lemma}

\begin{IEEEproof}
We have three type of \reference groups. We call a \reference group \emph{promising} if it contains at most $\lowt-1$ defective items, and call it \emph{misleading} if it contains at least $\lowt+1$ defective items. 
Finally, recall that a \reference group is critical if it contains exactly $\lowt$ defective items. The error events include four kinds of misclassification. We denote the probability of misclassifying a promising $\rfrn_{\dstr, \family}$ to be \critical~by $\pepvp$, the probability of misclassifying a \critical~$\rfrn_{\dstr, \family}$ to be promising by $\pevpp$, the probability of misclassifying a misleading $\rfrn_{\dstr, \family}$ to be {\critical} by $\pemvp$, and the probability of misclassifying a {\critical} $\rfrn_{\dstr, \family}$ to be misleading by $\pevpm$. Each of the error probabilities can be bounded by a binomial distribution. The first error event can be computed as
\begin{align}
\pepvp & \leq \prob \left ( \bigcup_{\dstr, \family: |\rfrn_{\dstr, \family} \cap \cD| < \lowt} \sum_\stage{\outptu{0}{\dstr, \family}{\stage}} \geq |\cidct^{(0)}|\rtrsd{\lowt}(1-\rtail{{<\lowt}}) \right ) \nonumber \\
& \leq \rrpt \bnum \sum_{t = \irpt \rtrsd{\lowt} (1-\rtail{{<\lowt}})}^{\irpt} \binom{\irpt}{t}\rtrsd{{\lowt-1}}^t(1-\rtrsd{{\lowt-1}})^{\irpt-t} \label{eq:pepvp1} \\
& \leq \rrpt \bnum~\text{exp} \left ( -2\irpt ( \rtrsd{\lowt}(1-\rtail{{<\lowt}}) - \rtrsd{{\lowt-1}} )^2 \right ). \label{eq:pepvp2}
\end{align}
In inequality (\ref{eq:pepvp1}), we take union bound over all $\rfrn_{\dstr, \family}$, and the summation is over the tail of a binomial distribution (corresponding to the event that a promising reference group ``behaves like'' a {\critical} \reference group).
Inequality (\ref{eq:pepvp2}) follows from the Chernoff bound. Similarly, for the other error events we have that
\begin{align}
\pevpp & \leq \rrpt \bnum~\text{exp} \left ( -2\irpt ( \rtrsd{\lowt}\rtail{{<\lowt}} )^2 \right ), \label{eq:pevpp2} \\
\pemvp & \leq \rrpt \bnum~\text{exp} \left ( -2\irpt ( \rtrsd{{\lowt+1}} - \rtrsd{\lowt}(1+\rtail{{>\lowt}}) )^2 \right ), \label{eq:pemvp2} \\
\pevpm & \leq \rrpt \bnum~\text{exp} \left ( -2\irpt ( \rtrsd{\lowt}\rtail{{>\lowt}} )^2 \right ). \label{eq:pevpm2}
\end{align}

Within the valid range of $\rtail{{<\lowt}}$, $\pepvp$ is strictly increasing as a function of $\rtail{{<\lowt}}$; conversely, $\pevpp$ is strictly increasing as a function of $\rtail{{<\lowt}}$. 
$\rtail{{<\lowt}}$ is one that allows for a ``small'' choice of $\lowt$, while still keeping both $\pepvp$ and $\pevpp$ ``small''. The same argument holds for $\rtail{{>\lowt}}$ to $\pemvp$ and $\pevpm$. Some specific choices of $\rtail{{<\lowt}}$ and $\rtail{{>\lowt}}$ that work, and that we use, are
\begin{align}
\rtail{{<\lowt}} = (\rtrsd{\lowt} - \rtrsd{{\lowt-1}})/(2\rtrsd{\lowt}), \label{eq:rtail1} \\
\rtail{{>\lowt}} = (\rtrsd{{\lowt+1}} - \rtrsd{\lowt})/(2\rtrsd{\lowt}), \label{eq:rtail2}
\end{align}
so that (\ref{eq:pepvp2}) balances (\ref{eq:pevpp2}), and (\ref{eq:pemvp2}) balances (\ref{eq:pevpm2}).

Let $\pis{w|\dsize} = \prob \left (| ( \idct^{(0)}_{\stage} \setminus \rfrn_{\dstr, \family} ) \cap \cD  | = w \bigm | |\rfrn_{\dstr, \family} \cap \cD| = \dsize \right )$ (This is the probability that, conditioned on the \reference group $\rfrn_{\dstr,\family}$ containing $\dsize$ items, a randomly chosen indicator set from the $\stage$-th family $ \idct^{(0)}_{\stage}$ contains exactly $w$ defective items that are not contained in the reference group $\rfrn_{\dstr,\family}$).

Hence the conditional probabilities of giving a positive outcome can be expanded as
\begin{align}
\rtrsd{{\lowt-1}} & = \frac{1}{2} \left ( 1 - \pis{0|\lowt-1} - \pis{1|\lowt-1} + \sum_{\dsize=\gap}^{\dfct} \pis{\dsize|\lowt-1} \right ) \label{eq:rtrsd1} \\
\rtrsd{{\lowt}} & = \frac{1}{2} \left ( 1- \pis{0|\lowt} + \sum_{\dsize=\gap}^{\dfct} \pis{\dsize|\lowt} \right ) \label{eq:rtrsd2} \\
\rtrsd{{\lowt+1}} & = \frac{1}{2} \left ( 1 + \sum_{\dsize=\gap}^{\dfct} \pis{\dsize|\lowt+1} \right ) \label{eq:rtrsd3}
\end{align}

Since the summations in each equation (\ref{eq:rtrsd1})-(\ref{eq:rtrsd3}) are ``close'' to each other, we ignore them in the following calculations (since only their pairwise differences required, and the summations only contribute lower-order terms). For example,
the difference between $\pis{\dsize|\lowt}$ and $\pis{\dsize|\lowt-1}$ can be computed as
\begin{align}
\pis{\dsize|\lowt} - \pis{\dsize|\lowt-1} & =  \pis{\dsize|\lowt} \left ( 1 - \frac{\pis{\dsize|\lowt-1}}{\pis{\dsize|\lowt}} \right ) \nonumber \\
& \doteq \pis{\dsize|\lowt} \left ( \frac{\dsize}{\dfct-\lowt+\dsize} \right ) \label{eq:posdiff}
\end{align}
When $\dsize=o(\dfct)$, (\ref{eq:posdiff}) is bounded from above by $\dsize/(\dfct-\lowt+\dsize)$, which is 
asymptotically negligible as $\dfct$ grows without bound.

The quantity $2\rtrsd{\lowt}\rtail{{>\lowt}}$ in (\ref{eq:pevpm2}) is then bounded by using (\ref{eq:rtail2}), (\ref{eq:rtrsd2}) and (\ref{eq:rtrsd3}), and noting that
\begin{align}
4\rtrsd{\lowt}\rtail{{>\lowt}} & = 2(\rtrsd{{\lowt+1}} - \rtrsd{\lowt}) \nonumber \\
& = \pis{0|\lowt} \nonumber \\
& = \prod_{i=0}^{\frac{\iratio \bl}{\dfct-\lowt}-1} \left ( 1 - \frac{\dfct-\lowt}{\bl - i} \right ) \nonumber \\
& > \left ( 1 - \frac{\dfct-\lowt}{\bl - \frac{\iratio \bl}{\dfct-\lowt}} \right )^{\frac{\iratio \bl}{\dfct-\lowt}} \nonumber \\
& \geq \text{exp} \left ( -\frac{\iratio\bl}{\bl - \frac{\iratio \bl}{\dfct-\lowt} - \dfct + \lowt} \right ) \label{eq:diffru3}
\end{align}
Inequality (\ref{eq:diffru3}) follows from the fact that $(1+x)^{-y} \geq e^{xy}$ for $|x|<1$ and $y \geq 0$. 

The quantity $2\rtrsd{\lowt}\rtail{{<\lowt}}$ in (\ref{eq:pevpp2}) can be bounded in a similar manner as
\begin{align}
4\rtrsd{\lowt}\rtail{{<\lowt}} & = 2(\rtrsd{{\lowt}} - \rtrsd{{\lowt-1}}) \nonumber \\
& > \pis{1|\lowt-1} \nonumber \\
& = \frac{ \binom{\dfct-\lowt+1}{1} \binom{\bl-\dfct+\lowt-1}{ \frac{\iratio\bl}{\dfct-\lowt} - 1 } }{ \binom{\bl}{ \frac{\iratio\bl}{\dfct-\lowt} } } \nonumber \\
& = \iratio \frac{\dfct-\lowt+1}{\dfct-\lowt} \prod_{i=0}^{\frac{\iratio \bl}{\dfct-\lowt}-2} \left ( 1 - \frac{\dfct - \lowt}{\bl - 1 - i} \right ) \nonumber \\
& > \iratio \left ( 1 - \frac{ \dfct - \lowt }{ \bl - \frac{\iratio \bl}{\dfct-\lowt} + 1 } \right )^{ \frac{\iratio \bl}{\dfct-\lowt} -1 } \nonumber \\
& \geq \iratio~\text{exp} \left ( -\frac{ \iratio\bl - \dfct + \lowt }{\bl - \frac{\iratio \bl}{\dfct-\lowt} + \dfct  - \lowt} \right ). \label{eq:diffrl3}
\end{align}

Finally we substitute the results from (\ref{eq:rtail1})-(\ref{eq:diffrl3}) into (\ref{eq:pepvp2})-(\ref{eq:pevpm2}). The requirement that error probability of misclassification of any \reference groups be at most $\error{3}$ implies
\begin{equation*}
\irpt > 8 \max \left ( \left ( \pis{0|\lowt} \right )^{-2} , \left ( \pis{1|\lowt-1} \right )^{-2} \right ) \left ( \ln(\rrpt \bnum) + \ln \left ( \frac{1}{\error{3}} \right ) \right ).
\end{equation*}
For $\dfct = o(\bl)$ and ``large enough'' $n$, (\ref{eq:diffru3}) ``behaves'' as $e^{\iratio}$, and (\ref{eq:diffrl3}) ``behaves'' as  $\iratio e^{-\iratio}$. The quantity $\irpt$ is minimized when $\iratio$ is 1. Therefore we obtain the result in Lemma~\ref{thm:decvp}.
\end{IEEEproof}

\begin{lemma}
With error probability at most $\error{4}$, the decoder correctly determines whether an item $\inp_\person$ is defective or non-defective when
\begin{equation*}
\irpt > 8 e^2 \left ( \ln(\bl) + \ln \left ( \frac{1}{\error{4}} \right ) \right ).
\end{equation*}
\label{thm:decinp}
\end{lemma}

\begin{IEEEproof}
As the rule for deciding whether an item is defective is not, and also the rule for deciding whether a \reference group is {\critical} or not, both depend on matching empirically observed test outcome statistics with precomputed thresholds, the proof here is essentially the same as in the proof of Lemma~\ref{thm:decvp}. We outline the major changes below.

The error event includes both false positives (misclassifying non-defective items to be defective) and false negatives (misclassifying defective items to be non-defective).The probability of false negatives can be computed as
\begin{align}
P_e^- & = \prob \left ( \bigcup_{\person : \inp_\person = 1} \sum_\stage{\outptu{\person}{\family}{\stage}} \leq |\cidct^{(\person)}|\itrsd{0}(1+\itail) \right ) \nonumber \\
& < \bl \sum_{t = \irpt \itrsd{0} (1-\itail)}^{\irpt} \binom{\irpt}{t} (1-\itrsd{1})^{t} \itrsd{1}^{\irpt-t} \nonumber \\
& \leq \bl~\text{exp} \left ( -2\irpt ( \itrsd{1} - \itrsd{0}(1-\itail) )^2 \right ).
\label{eq:falsepos2}
\end{align}
In a similar manner, the probability of false positives can be computed as
\begin{align}
P_e^+ & = \prob \left ( \bigcup_{\person : \inp_\person = 0} \sum_\stage{\outptu{\person}{\family}{\stage}} > |\cidct^{(\person)}|\itrsd{0}(1+\itail) \right ) \nonumber \\
& < \bl \sum_{t = \irpt \itrsd{0} (1-\itail)}^{\irpt} \binom{\irpt}{t} \itrsd{0}^{t} (1-\itrsd{0})^{\irpt-t}\nonumber \\
& \leq \bl~\text{exp} \left ( -2\irpt ( \itrsd{0}\itail )^2 \right ). \label{eq:falseneg2}
\end{align}

Let $\pnd{\dsize} = \prob(| \idct^{(\person)}_{\stage} \cap \cD  | = \dsize \bigm | \inp_\person = 0)$ and $\pd{\dsize} = \prob(| \idct^{(\person)}_{\stage} \cap \cD  | = \dsize \bigm | \inp_\person = 1)$. A good choice of $\itail$ is
\begin{equation}
\itail = ( \itrsd{1} - \itrsd{0} ) / (2\itrsd{0}). \label{eq:itail}
\end{equation}
We may expand $\itrsd{0}$ and $\itrsd{1}$ in terms of $\pnd{\dsize}$ and $\pd{\dsize}$ as
\begin{align*}
\itrsd{0} & = \frac{1}{2} \left ( 1 - \pnd{0} + \sum_{\dsize=\gap}^{\dfct} \pnd{\dsize} \right ) \\
\itrsd{1} & = \frac{1}{2} \left ( 1 + \sum_{\dsize=\gap}^{\dfct} \pd{\dsize} \right ).
\end{align*}
The difference between $\itrsd{0}$ and $\itrsd{1}$ can be computed as
\begin{align}
2(\itrsd{1}-\itrsd{0}) & > \pnd{0} \label{eq:pnd0} \\
& = \prod_{i=0}^{\frac{\iratio \bl}{\dfct-\lowt}-2} \left ( 1 - \frac{\dfct - \lowt}{\bl - 1 - i} \right ) \nonumber \\
& > \left ( 1 - \frac{ \dfct - \lowt }{ \bl - \frac{\iratio \bl}{\dfct-\lowt} + 1 } \right )^{ \frac{\iratio \bl}{\dfct-\lowt} -1 } \nonumber \\
& \geq \text{exp} \left ( -\frac{ \iratio\bl - \dfct + \lowt }{\bl - \frac{\iratio \bl}{\dfct-\lowt} - \dfct + \lowt} \right )
\label{eq:pnd3}
\end{align}
Equality (\ref{eq:pnd0}) ignores the small difference came from the summation terms. Inequality (\ref{eq:pnd3}) follows from the fact that $(1+x)^{-y} \geq e^{-xy}$ for $|x|<1$ and $y \geq 0$.

FInally we substitute (\ref{eq:itail}) and (\ref{eq:pnd3}) into (\ref{eq:falsepos2}) and (\ref{eq:falseneg2}), and $\iratio$ is set to be $1$ in Lemma~\ref{thm:decvp}. The requirement that the error probability of misclassification of any item be at most $\error{4}$ implies the result in Lemma~\ref{thm:decinp}.
\end{IEEEproof}

\noindent {\bf Proof of Theorem~\ref{thm:tgt1}:}
A sufficient condition for high probability decoding all items is when Lemmas~\ref{thm:allref}-\ref{thm:decinp} are satisfied. Therefore, with error probability at most $\error{2}+\error{3}+\error{4}$, the total number of tests $\test$ is  $\rrpt \bnum (\dfct-\lowt) \irpt$, where $\bnum=(2\dfct)/(\dfct-\lowt)$, $\rrpt$ is specified in Lemma~\ref{thm:allref}, and $\irpt$ required to satisfy both Lemma~\ref{thm:decvp} and Lemma~\ref{thm:decinp} is set as
\begin{equation*}
\irpt > 8 e^2 \left ( \ln(\bl) + \ln \left ( \frac{1}{\error{3}} \right ) + \ln \left ( \frac{1}{\error{4}} \right ) \right ).
\end{equation*}
Explicitly, $\test$ is at least $(4e^8\ln(2)/\pi^2)\ln(1/\error{2})\sqrt{\lowt}\dfct\ln(\bl) + \bigo{\ln(1/\error{3}\error{4})\dfct\sqrt{\lowt}}$. As to the computational complexity of decoding, recall that the first decoding step decodes a \reference group by counting the empirical fraction of positive outcomes from $\irpt$ \indicator groups, and the second decoding step decodes an item by doing the same thing. Therefore, given there are $\bnum\rrpt$ \reference groups and $\bl$ items, the complexity is $\irpt(\bl+\bnum\rrpt)$, which is $\bigo{\bl \ln(\bl)} + \bigo{\ln(1/\error{2})\ln(\bl)} + \bigo{\bl\ln(1/\error{2}\error{3})}$. Let $\error{}=\max(\error{2}, \error{3}, \error{4})$, we obtain Theorem~\ref{thm:tgt1}.
\hfill  $\blacksquare$

\section{Proof sketches of Theorems~\ref{thm:tgt2} and~\ref{thm:tgt3}}
Slight modifications of {\algoa} can result in an adaptive algorithm (as in Theorem~\ref{thm:tgt2}), and also an algorithm for threshold group testing models where the probability of giving a positive outcome is a monotonically nondecreasing function of the number of defective items being measured. As a demonstration, we show a two-stage adaptive algorithm {\algob}, and a non-adaptive algorithm {\algoc} that works under the ``linear model'' of stochastic threshold group testing (as in Figure~\ref{fig:model2}).

\subsection{\algob}
In the first stage, we aim to find multiple {\critical} \reference groups. This is done by first performing the encoding step 1 of {\algoa}, and then obtaining a set of \indicator groups called $\cidct^{(0)}$. The construction of $\cidct^{(0)}$ is however slightly different from the definition given in {\algoa}. Here $\cidct^{(0)}=\{ \idct^{(0)}_\stage : \stage = 1,2,\cdots, \irpt_1 \}$, where each $\idct^{(0)}_\stage$ is a group of $\iratio\bl/\dfct$ distinct items randomly picked from $\cN$. For every pair of a \reference group from $\crfrn$ and an \indicator group from $\cidct^{(0)}$, the two groups are pooled together and a threshold group test is performed.  As to inference of whether particular \reference groups are critical or not, this follows the decoding step 1 of {\algoa}, but using the definition/parameters of $\cidct^{(0)}$ provided in this paragraph.

The second stage follows steps 2 and 3 of {\algoa}. However, only the set of \reference groups decoded to be critical in the first stage are tested in step 3, hence the multiplicative factor of $\sqrt{l}$ is missing from the overall number of tests. To avoid confusion with {\algoa}, notation for the total number of families in the set of \indicator groups (denoted $\irpt$ in {\algoa}) is replaced by $\irpt_2$. Finally, to decode whether individual items are defective or not, {\algob} uses decoding step 2 in {\algoa}.


\noindent {\bf Proof sketch of Theorem~\ref{thm:tgt2}:}
A sufficient condition for high probability decoding all items is when $\rrpt$ satisfies Lemma~\ref{thm:allref}, $\irpt_1$ satisfies Lemma~\ref{thm:decvp}, and $\irpt_2$ satisfies Lemma~\ref{thm:decinp}. Therefore, with error probability at most $\error{2}+\error{3}+\error{4}$, the total number of tests $\test$ is $\rrpt \bnum \irpt_1 + \bnum (\dfct-\lowt) \irpt_2$, which is at least $16e^2\dfct\ln(\bl) + \bigo{\ln(1/\error{2})\sqrt{\lowt}\ln(\lowt)} + \bigo{\ln(1/\error{3})\sqrt{\lowt}} + \bigo{\ln(1/\error{4})\dfct}$. The computational complexity of decoding is $\bl\irpt_2+\bnum\rrpt\irpt_1$, which is $\bigo{\bl\ln(\bl)} + \bigo{\ln(1/\error{2})\sqrt{\lowt}\ln(\lowt)} + \bigo{\ln(1/\error{3})\sqrt{\lowt}} + \bigo{\ln(1/\error{4})\bl} $. Setting $\error{}=\max(\error{2}, \error{3}, \error{4})$, we obtain Theorem~\ref{thm:tgt2}.

\hfill  $\blacksquare$

\subsection{\algoc}
The testing scheme follows that of {\algoa} exactly. The decoding scheme is based on {\algoa}, but has the following changes:
\begin{enumerate}
\item{
\underline{Estimation of number of defectives in a reference group:}
We first estimate the number of defective items in a {\it single} reference group, according to the empirical probability the reference group resulting in positive test outcomes. More precisely,
let $\rtrsd{\dsize} = \prob \left ( \outptu{0}{\dstr, \family}{\stage} = 1 \bigm | |\rfrn_{\dstr, \family} \cap \cD| = \dsize \right )$ be the expected fraction of positive test outcomes. The decoder declares $\rfrn_{\dstr, \family}$ contains $\dsize$ defective items if
\begin{align*}
|\cidct^{(0)}|\rtrsd{\dsize}(1-\rtail{{<\dsize}}) \leq \sum_\stage{\outptu{0}{\dstr, \family}{\stage}} \leq |\cidct^{(0)}|\rtrsd{\dsize}(1+\rtail{{>\dsize}}),
\end{align*}
where the ``variation''  $\rtail{{<\dsize}}$ around the expectation is set to equal $(\rtrsd{\dsize} - \rtrsd{{\dsize-1}})/(2\rtrsd{\dsize})$ and $\rtail{{>\dsize}} = (\rtrsd{{\dsize+1}} - \rtrsd{\dsize})/(2\rtrsd{\dsize})$.
}
\item{
\underline{Estimation of defectiveness of items:} In this case, the threshold for estimating that an item is defective is different than in \algoa, since the variation between the empirical probability of observing positive testing outcomes in tests including the reference group is different.
Specifically, let $\itrsd{w,\dsize} = \prob \left ( \outptu{\person}{\dstr, \family}{\stage} = 1 \bigm | |\rfrn_{\dstr, \family} \cap \cD| = \dsize, \inp_{\person} = \ans \right )$. For every $\rfrn_{\dstr, \family} \subseteq \bin_\dstr$ which contains $\dsize \in (\lowt, \upt)$ defective items and $\inp_\person \in \parta_\dstr$, if $\sum_\stage{\outptu{\person}{\dstr, \family}{\stage}} \leq |\cidct^{(\person)}|\itrsd{\dsize,0}(1+\itail)$ where $\itail_\dsize = ( \itrsd{\dsize,1} - \itrsd{\dsize,0} ) / (2\itrsd{\dsize,0})$, the decoder declares $\inp_j$ to be non-defective, else declares it to be defective.
}
\end{enumerate}


\noindent {\bf Proof sketch of Theorem~\ref{thm:tgt3}:}
We note that Lemma~\ref{thm:allref} is not required, since any \reference groups $\rfrn$ can be used to decode items, as long as the number of defective items in $\rfrn$ is between $\lowt$ and $\upt$, so that there is some statistical difference between the probability of a positive test outcome if a group has $i$ defective items, or if it has $i+1$ defective items). With high probability, a \reference group with a suitably chosen size ($n(u+l)/2d$) satisfies this relaxed condition. That is, the number of reference groups $\rrpt$ required is a constant.

A sufficient condition of high probability decoding of all items is when modified versions of Lemmas~\ref{thm:decvp} and~\ref{thm:decinp} are satisfied. The modifications in Lemma~\ref{thm:decvp} and~\ref{thm:decinp} correspond to the fact that the required number of \indicator groups increase by a factor of $\gap^2$. This is because the decoder is based on estimating the probability difference of giving a positive outcome when the test has an additional defective item. Hence the difference between two different probabilities scales as $1/\gap$. By the Chernoff bound, to estimate such a probability difference sufficiently accurately requires a multiplicative factor of $\gap^2$ in the number of tests.

The rest of the argument is as in the proof of Theorem~\ref{thm:tgt1}.

\hfill  $\blacksquare$

\bibliographystyle{IEEEtran}
\bibliography{tgt}

\end{document}